\documentclass[
  aps,
  prl,
  reprint,
  superscriptaddress,
  amsmath,
  amssymb,
  longbibliography]{revtex4-2}

\usepackage{graphicx}
\usepackage{dcolumn}
\usepackage{bm}
\usepackage{siunitx}
\usepackage[noend]{algorithmic}
\usepackage{algorithm}
\usepackage{setspace}
\usepackage{blkarray}

\usepackage{xcolor}
\usepackage{soul}
\usepackage{systeme}
\usepackage{footmisc}
\usepackage{mhchem}
\usepackage{braket}
\usepackage{comment}
\usepackage{float}

\usepackage{scalerel} 

\makeatletter
\def\maketag@@@#1{\hbox{\m@th\normalfont\normalsize#1}}
\makeatother

\newcommand{\bs}{\boldsymbol}

\usepackage{booktabs}
\usepackage{hyperref}
\usepackage[capitalize]{cleveref}
\usepackage[utf8]{inputenc}

\usepackage[lining,semibold]{libertine} 
\usepackage{amsthm}
\usepackage[libertine, cmintegrals, bigdelims, vvarbb]{newtxmath}

\usepackage{amsmath}
\usepackage{amsfonts}
\usepackage{mathrsfs}
\usepackage{gensymb}
\usepackage{bbm}
\usepackage{dsfont}

\usepackage{pgfplots}
\usepgfplotslibrary{colormaps}

\usepackage{chemformula}
\usepackage[caption=false]{subfig}

\usepackage{soul}
\usepackage{xcolor}
\usepackage{enumitem}

\theoremstyle{definition}

\usepackage{scalerel} 

\definecolor{webgreen}{rgb}{0,.5,0}
\definecolor{webbrown}{rgb}{.6,0,0}
\definecolor{grigio}{rgb}{.85,.85,.85} 
\definecolor{RoyalBlue}{rgb}{0.0, 0.14, 0.4}
\definecolor{skyblue1}{rgb}{0.45,0.62,0.81}
\definecolor{skyblue2}{rgb}{0.2,0.39,0.64}
\definecolor{skyblue3}{rgb}{0.13,0.29,0.53}
\definecolor{scarlet1}{rgb}{0.93,0.16,0.16}
\definecolor{scarlet2}{rgb}{0.8,0,0}
\definecolor{scarlet3}{rgb}{0.64,0,0}

\definecolor{g}{gray}{0.50}

\hypersetup{%
    colorlinks=true, linktocpage=true, pdfstartpage=1, pdfstartview=FitV,%
    breaklinks=true, pdfpagemode=UseNone, pageanchor=true, pdfpagemode=UseOutlines,%
    plainpages=false, bookmarksnumbered, bookmarksopen=true, bookmarksopenlevel=1,%
    hypertexnames=true, pdfhighlight=/O,
    urlcolor=webbrown, linkcolor=RoyalBlue, citecolor=webgreen, 
    pdftitle={},%
    pdfauthor={Norayr Asriyan},%
    pdfsubject={},%
    pdfkeywords={},%
    pdfcreator={pdfLaTeX},%
    pdfproducer={LaTeX REVTeX}%
}

\begin{document}
%
\title{Dynamical phase transitions for single particles in the  semiclassical and weak noise limits}
\author{Norayr Asriyan}
\email{norair.asriian@uni.lu}
\affiliation{Complex Systems and Statistical Mechanics, Department of Physics and Materials Science, University of Luxembourg, 30 Avenue des Hauts-Fourneaux, L-4362 Esch-sur-Alzette, Luxembourg}
\author{Jan Meibohm}
\affiliation{Theoretical Physics Unit, Institute for Physics and Astronomy, Technische Universität Berlin, Hardenbergstraße 36, 10623 Berlin, Germany}
\author{Vasco Cavina}
\affiliation{Scuola Normale Superiore, 56126 Pisa, Italy}
\author{Massimiliano Esposito}
\affiliation{Complex Systems and Statistical Mechanics, Department of Physics and Materials Science, University of Luxembourg, 30 Avenue des Hauts-Fourneaux, L-4362 Esch-sur-Alzette, Luxembourg}
\date{\today}

\begin{abstract}
We present a unifying description of dynamical phase transitions in the unitary evolution of an isolated quantum particle and the dissipative relaxation of a classical Brownian particle, based on a dynamical generating function. In the semiclassical and weak-noise limits, Fisher zeros of this function condense in the complex-time plane and reach the corresponding physical axes, producing dynamical phase transitions through a competition between return trajectories. This establishes a direct connection between quantum and classical (finite-time) dynamical phase transitions, where the semiclassical limit plays the role of the thermodynamic limit.  In particular, the established link naturally provides a classical version of the Loschmidt amplitude and shows how dynamical quantum phase transitions, typically associated with isolated many-body systems, can arise in a single-particle quantum system. The dynamical phases are distinguished by a unifying, trajectory-based order parameter, realized as a classical correlation function and a sequential quantum weak value.
\end{abstract}

\maketitle

\paragraph{Introduction ---}  
Equilibrium phase transitions emerge in the thermodynamic limit, when the zeros of the partition function condense into curves that pinch the real axis in the complexified control-parameter plane~\cite{Fisher1965Nature, PhysRev.87.404, PhysRev.87.410}. This produces nonanalyticities in thermodynamic potentials, while the different phases are commonly distinguished by an order parameter~\cite{landau1980, huang1987}.  

Dynamical quantum phase transitions (DQPTs) extend this framework to nonequilibrium quantum dynamics~\cite{PhysRevLett.110.135704, heylDynamicalQuantumPhase2018}.
In this setting, a system initially prepared in the state $\ket{\psi_0}$ is quenched and subsequently evolves in time. The survival amplitude (SA), $\braket{\psi_0|\psi(t)}$, also known as the Loschmidt amplitude, then plays the role of a dynamical partition function and satisfies a large-deviation form in the thermodynamic limit $N\to\infty$. Time thus naturally assumes the role of the control parameter, while Fisher zeros of the SA may approach the real-time axis in this limit, giving rise to nonanalyticities in its rate function at critical times.
A less universal aspect of the analogy is the characterization of the dynamical phases by an order parameter. Dynamical order parameters have been identified for various classes of DQPTs, including quantities inherited from the underlying equilibrium transition and genuinely dynamical topological order parameters~\cite{PhysRevLett.113.205701,BudichHeyl2016,heylDynamicalQuantumPhase2018}. However, no universal construction of an order parameter for arbitrary DQPTs is known, and the existence of a local dynamical order parameter generally depends on the model~\cite{Kuliashov2023}.

A complementary perspective is offered by a recently identified classical \textit{finite-time} dynamical phase transition occurring during the nonequilibrium relaxation of the Curie-Weiss model following a quench from the ferromagnetic to the paramagnetic phase~\cite{meibohmFiniteTimeDynamicalPhase2022}. In this setting, a nonanalyticity emerges in the magnetization distribution from the competition between optimal relaxation trajectories. The initial magnetization of the optimal trajectory provides a natural dynamical order parameter whose behavior closely parallels that of its equilibrium counterpart, including the same mean-field critical exponents. Subsequent work has extended this framework to thermodynamic observables~\cite{Meibohm_2023}, short-range interacting systems~\cite{Vadakkayil2024,blom2023globalspeedlimitfinitetime}, and optimal control~\cite{Meibohm_2026}, revealing nontrivial kinetic constraints and critical fluctuations. Yet a key element of the equilibrium analogy remains missing: unlike for DQPTs, no counterpart of the dynamical partition function has been identified whose rate function becomes nonanalytic in time. This raises a first fundamental question: are these classical and quantum dynamical phase transitions manifestations of the same underlying mechanism, and can they be described within a unified framework? Such a framework should, in particular, provide a classical counterpart of the dynamical partition function and its Fisher zeros. 

Moreover, whereas DQPTs emerge in the conventional thermodynamic limit $N\to\infty$, the relaxation transition described above points to a broader notion of this limit. In particular, Refs.~\cite{meibohmFiniteTimeDynamicalPhase2022,Meibohm_2026} show that this transition is not inherently a many-body phenomenon: a mathematically equivalent transition occurs in the weak-noise relaxation of a single Brownian particle, with the inverse noise strength playing the role of system size. 
This raises a second fundamental question: can an analogous DQPT arise in an isolated single-particle quantum system in the semiclassical limit $\hbar\to 0$, with $\hbar$ playing a role analogous to that of the noise strength?

In this Letter, we construct quantum and classical single-particle problems that answer these questions within a unified framework. We first consider a quantum particle whose pre-quench state comprises two symmetry-related semiclassical sectors. Following a quench of the potential, competition between trajectories connecting distinct symmetry sectors generates nonanalyticities in the corresponding rate function in the semiclassical limit, closely paralleling many-body DQPTs involving symmetry-broken ground states~\cite{PhysRevLett.113.205701}.

We then show that, upon analytic continuation to imaginary time, the SA of the quantum problem maps onto a weighted return probability of a Brownian relaxation process, with the inverse noise strength playing the role of the large parameter. The quantum and classical problems thus emerge as different cuts through a common complex-time phase diagram and share the same Fisher-zero structure. Finally, we identify a common dynamical order parameter for both problems and compare the transition introduced here with that of Ref.~\cite{meibohmFiniteTimeDynamicalPhase2022}.

\paragraph{Quenched dynamics of a quantum particle ---} We consider a quantum particle of mass $m$ initialized in a simple double-peaked state
\begin{align}\label{eq:double-peak}
	\psi_0(x)\sim \exp\left[-\frac{\alpha m\omega a^2}{2\hbar}\left(\frac{x^2}{a^2}-1\right)^2\right]\,,
\end{align}
where $\alpha a^2$ controls the degree of localization relative to $\ell^2 = \hbar/m\omega$.
In 1D such a smooth nodeless state is always a ground state of a suitably constructed potential (see Appendix \ref{app:SUSY}). At $t=0$, the system undergoes a sudden quench into a harmonic well $V(x)=(m\omega^2x^2-\hbar\omega)/2$.   With dimensionless variables $\xi=x/a$, $\tau_x = \omega t$, the post-quench Schr\"odinger equation reads
\begin{align}\label{eq:dimensionless_Shrodinger}
	i\partial_{\tau_x}\psi(\xi,\tau_x)=-\frac{\eta}{2}\partial_\xi^2\psi(\xi,\tau_x)+\left(\frac{\xi^2}{2\eta}-\frac{1}2\right)\psi(\xi,\tau_x) \;,
\end{align}
with an initial condition $\psi_0(\xi)\sim \exp\left[-\frac{\alpha}{2\eta}(\xi^2-1)^2\right]$. The parameter $\eta=\hbar/(m\omega a^2)$ governs the overall degree of semiclassical localization.

\paragraph{Survival rate function for the quantum particle ---} The signature of a DQPT is the nonanalytic behavior of the survival rate function (SRF)
\begin{align}\label{eq:SRF_def}
	g(\tau_x) = \lim_{\eta\to 0} \left\{-\eta \ln[|G(\tau_x)|]\right\},
\end{align}
where $G(\tau_x){=}\bra{\psi_0}e^{-i\hat H \tau_x/(\hbar\omega)}\ket{\psi_0}$ denotes the SA.
Here $\eta$ plays the role of an inverse system size in the semiclassical limit. In terms of the post-quench propagator $K(\xi, \xi_0, \tau_x)$:
\begin{align}\label{eq:SRF_integral}
	&G(\tau_x){=}\int\limits_{-\infty}^{\infty}\!\!\!d\xi \psi_0(\xi)\!\!\int\limits_{-\infty}^{\infty}\!\!\!d\xi_0K(\xi,\xi_0, \tau_x)\psi_0(\xi_0).
\end{align}
\begin{figure}
\includegraphics[
trim = 2mm 4mm 2mm 10mm,
clip,
width=\linewidth]{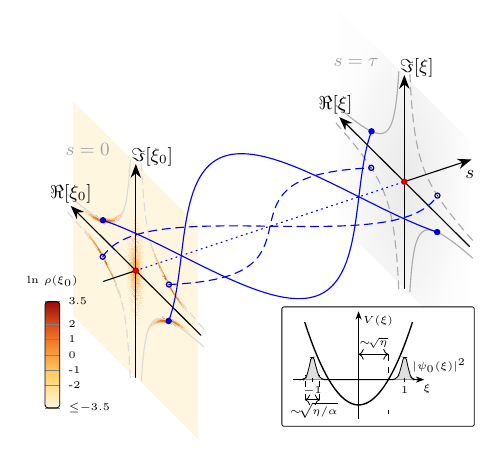}
\caption{Trajectories corresponding to the saddles of $S$. Solid blue lines show a pair of ''direct'' trajectories ($\xi = \xi_0$), dashed blue lines show the ''exchange'' trajectories ($\xi = -\xi_0$). The ''origin-returning'' trajectory $0\to0$ is the dotted blue line. The solid and dashed gray curves on the $s = 0$ and $s=\tau$ planes are the projections of the steepest-descent surface onto the $\xi=\xi_0$ and $\xi = -\xi_0$ subspaces respectively. The density plot on the $s=0$ plane shows the numerically sampled initial-endpoint density $\rho(\xi_0)$. Here $\eta=0.3$, $\alpha = 0.65$ and $\tau_x = 4.75$ (for this particular time the direct saddle is dominant). The density plot is a result of numerical sampling of $N=10^6$ trajectories (see details in Appendix~\ref{app:PL}). The inset shows the post-quench harmonic potential and the initial probability distribution.}
\label{fig:3Dtraj}
\end{figure}

The harmonic oscillator propagator is given by the Mehler kernel~\cite{10.1119/1.1538573}, allowing us to write
\begin{align}
	G(\tau_x)\sim \int d\xi d\xi_0 \exp{\left(-\frac{S(\xi, \xi_0, \tau_x)}{\eta}\right)}
\end{align}
with
\begin{align}\label{eq:action}
	\!\!\!S(\xi, \xi_0, \tau_x){=}\frac{\alpha(\xi^2{-}1)^2}2{+}\frac{\alpha(\xi_0^2{-}1)^2}2{+}\frac{(\xi^2{+}\xi_0^2)\cos(\tau_x){-}2\xi\xi_0}{2i\sin(\tau_x)}.
\end{align}
Semiclassically, the propagator is dominated by fixed-endpoint classical paths,
while the endpoints are weighted by the initial wave function. The dominant contributions are determined by the nine saddles of $S$, which generically fall into four classes labeled by $\sigma\in\{d,e,0,m\}$: two direct saddles ``$d$'' with $\xi_0=\xi$, two exchange saddles ``$e$'' with $\xi_0=-\xi$, the saddle at the origin ``$0$'' at $\xi_0=\xi=0$, and four mixed saddles ``$m$'' with $\xi \ne \pm \xi_0$. Each saddle specifies the endpoints of the following classical trajectory
\begin{align}
	\xi(s)=\frac{\xi_0 \sin(\tau_x-s)+\xi\sin(s)}{\sin(\tau_x)}, \quad s\in [0;\tau_x].
\end{align}
Gaussian expansion around the saddles~\cite{pemantle2009asymptoticexpansionsoscillatoryintegrals} gives
\begin{align}\label{eq:saddle_expansion}
G(\tau_x)\simeq	\sum_{\sigma\in\{d,e,0,m\}}g_\sigma n_\sigma\frac{2\pi\eta}{\sqrt{\det H_\sigma}}e^{-S_\sigma(\tau_x)/\eta},
\end{align}
where \((g_d,g_e,g_0,g_m)=(2,2,1,4)\) are the degeneracies, while \(n_\sigma\in\{0,1\}\) indicates whether the original integration plane can be deformed through the corresponding saddle along steepest-descent directions. Determining \(n_\sigma\) is, in general, a nontrivial global problem, a subject of the Picard--Lefschetz theory~\cite{witten2010analyticcontinuationchernsimonstheory,TANIZAKI2014250}. In Appendix~\ref{app:PL}, analytic continuation determines $n_\sigma$ and shows that mixed saddles never contribute to the integral~\eqref{eq:SRF_integral} as $\eta\to0$. Figure~\ref{fig:3Dtraj} illustrates the saddle trajectories and their
finite-$\eta$ endpoint broadening.

\begin{figure}
\includegraphics[width = \linewidth]{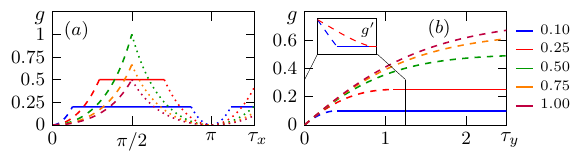}
\caption{The SRF for the quantum ($a$) and classical ($b$) problems.
Dashed, solid, and dotted lines indicate dominance of the direct,
origin, and exchange saddles, respectively. Kinks occur in $g$
for the quantum evolution and in $g'$ (inset) for the classical relaxation.}
\label{fig:SRF}
\end{figure}

Among the contributing saddles, the leading exponent is set by $\min\limits_{\sigma}\Re[S_\sigma]$. Importantly, as $\tau_x$ varies, the dominant contribution switches among the direct, exchange, and origin saddles; see also Fig. \ref{fig:SRF} ($a$): 
\begin{equation}
\label{eq:quantum_SRF}
	g(\tau_x){=}\min\left\{\frac{1}{4\alpha}\tan^2\left(\frac{\tau_x}2\right), \frac{1}{4\alpha}\cot^2\left(\frac{\tau_x}2\right), \alpha\right\}.
\end{equation}
The critical times are $\tau^{*}_{\pm}=\pm 2{\arctan}(2\alpha)+\pi n$ for $\alpha<1/2$ and $\tau^*=\pi/2+\pi n$ for $\alpha\geq 1/2$. At these times the switch of dominance is marked by a jump in the first derivative of the SRF and the phase transition is, therefore, of first order.

With this example, we demonstrate that a single-particle system can exhibit a DQPT. Unlike previously reported single-particle cusps in the survival probability~\cite{Zhang_2016, Yang_2019, Yang_2019_math}, the singularity here emerges in a large-deviation rate function in the controlled semiclassical limit $\eta\to 0$, as in conventional many-body DQPTs.

\paragraph{Quenched dynamics of a Brownian particle ---} Recalling the correspondence between the Fokker-Planck equation and an effective Schrödinger equation in imaginary time \cite{risken_fokker-planck_1996}, we construct a Fokker-Planck equation for which \eqref{eq:dimensionless_Shrodinger} is the effective imaginary-time equation with 
$\tau = -i\tau_y$. The transformation $P(\xi, \tau_y){=}\sqrt{P_{\rm eq}(\xi)}\psi(\xi, \tau_y)$ gives:
\begin{align}\label{eq:FP}
	\partial_{\tau_y} P(\xi, \tau_y)=\partial_\xi\left[\xi P(\xi, \tau_y)\right]+\frac{\eta}{2}\partial^2_{\xi}P(\xi, \tau_y)
\end{align}
with $P_{\rm eq}(\xi)=(\pi\eta)^{-1/2}\exp\left(-\xi^2/\eta\right)$ being the stationary distribution.
\begin{figure}[htp]
	\includegraphics[width = 0.9\linewidth,
  trim=0mm 0mm 0mm 0mm,
  clip]{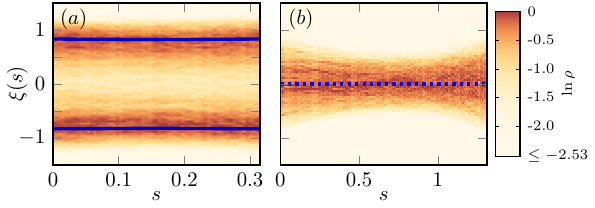}
	    \caption{Trajectory density from $N=10^8$ Brownian trajectories for
$\eta=0.1$, $\alpha=0.25$, and (a) $\tau_y=0.3$; (b) $\tau_y=1.3$,
before and after the phase transition, respectively. As in Fig.~\ref{fig:3Dtraj}, lines show the direct (solid blue) and origin (dotted black) saddles. The exchange saddle does not contribute at these times (see Fig. \ref{fig:statfiz} (a) in Appendix \ref{app:PL}).}
\label{fig:2Dtraj}
\end{figure}
The corresponding Wick-rotated SRF then reads 
\begin{align}\label{eq:g_FP}
	g(-i\tau_y)=\lim_{\eta\to 0} \left(-\eta \ln\left[\int_{-\infty}^{\infty}d\xi \frac{P_0(\xi)P(\xi, \tau_y)}{P_{\rm eq}(\xi)}\right]\right).
\end{align}
The weighting in \eqref{eq:g_FP} is not ad hoc: the quantity in brackets is
precisely the imaginary-time overlap $\langle\psi_0|\psi(\tau_y)\rangle$,
and thus a natural classical counterpart of the survival amplitude.

Equation~\eqref{eq:FP} is precisely the Fokker-Planck equation for an overdamped Brownian particle of mass $m$ and mobility $\mu$ in the harmonic potential $U(x)=m\omega^2x^2/2$, exposed to an Ohmic bath at temperature $T$. The correspondence follows from the dimensionless variables $\xi=x/a$, $\tau_y=m\omega^2\mu t$, and $\eta=2kT/(m\omega^2a^2)$. Furthermore, if the particle is initialized in the stationary state of the potential 
\begin{align}\label{eq:classical_prequench}
U_0=\frac{\alpha m\omega^2(x^2-a^2)^2}{4a^2}+\frac{m\omega^2 x^2}{4},
\end{align}
not only will the dynamical equations be the same for the quantum and classical problems (up to a Wick rotation), but the equivalence will also extend to \eqref{eq:SRF_def} and \eqref{eq:g_FP}.
\paragraph{Survival rate function for the classical Brownian particle ---} Continuing Eq. \eqref{eq:saddle_expansion} to imaginary time, $\tau_x\to-i\tau_y$, we identify the dominant saddles and evaluate the SRF for $\eta\to0$, which we recognize as the weak-noise limit [Fig.~\ref{fig:SRF}(b)]:
\begin{equation}\label{eq:classical_SRF}
	g(-i\tau_y)=\begin{cases}
		\tanh\left(\frac{\tau_y}{2}\right)\left[1-\frac{1}{4\alpha}\tanh\left(\frac{\tau_y}{2}\right)\right],\; \tau_y<\tau_y^*,\\
		\alpha,\; \tau_y\geq\tau_y^*.
	\end{cases}
\end{equation}
Now the exchange saddle never dominates, leaving a single nonanalyticity at $\tau^{*}_y=2{\rm arctanh}(2\alpha)$, where the dominant contribution switches from the direct to the origin-returning saddle. The transition exists only for $\alpha<\frac12$. 
\begin{figure}
    \hspace*{-0.4cm}%
\includegraphics[
  width=\linewidth,
  trim=0mm 6mm 0mm 4mm,
  clip
]{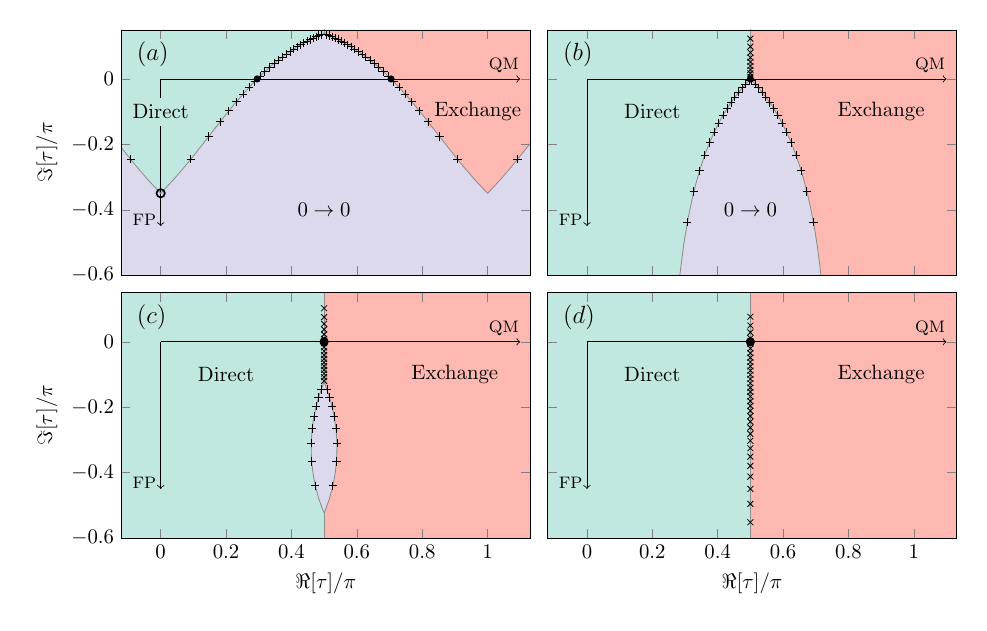}
\caption{Dynamical phase diagram and zeros of $G(\tau)$ for $\eta=0.01$: (a) $\alpha=0.25$; (b) $\alpha=0.5$; (c) $\alpha=0.65$; (d) $\alpha=0.8$. Zeros of $G(\tau)$ are shown as scatter points. Filled and hollow circles mark first- and second-order phase transition points, respectively.}
    \label{fig:Fisher_zeros}
\end{figure}

\begin{figure*}[t]
	\includegraphics[width = \linewidth,
  trim=0mm 8mm 0mm 2mm,
  clip]{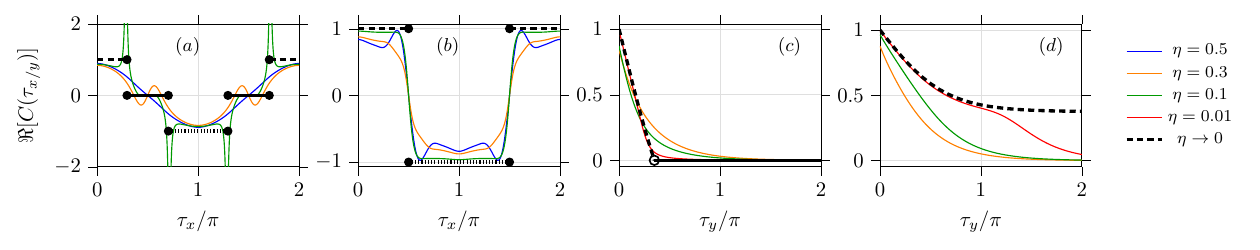}
	    \caption{Order parameter time dependence for different values of $\eta$, in the quantum case for (a) $\alpha = 0.25$ and (b) $\alpha = 0.8$; and in the classical case for (c) $\alpha = 0.25$ and (d)  $\alpha = 0.8$. Filled and hollow points mark phase transitions in accordance with panels (a) and (d) in Fig.~\ref{fig:Fisher_zeros}. Thick black lines show the asymptotic behavior for $\eta\to 0$.}
\label{fig:order_parameter}
\end{figure*}

As a result, the SRF and its first derivative are continuous, whereas the second derivative has a jump, indicating a second-order phase transition. Figure~\ref{fig:2Dtraj} provides the classical counterpart of the trajectory visualization in Fig. \ref{fig:3Dtraj}, now using numerically sampled stochastic trajectories on the two sides of the transition. Unlike in the quantum case, where we show only the endpoint distribution, here the full trajectory density can be displayed directly.

\paragraph{Dynamical phase transitions from Fisher zeros of $G(\tau)$ ---} 

The two problems we constructed are unified by introducing the complex time $\tau=\tau_x-i\tau_y$ and locating the zeros of $G(\tau)$. The corresponding analytical treatment is given in Appendices~\ref{app:PL} and~\ref{app:zeros}; Fig. \ref{fig:Fisher_zeros} displays the zeros for different values of $\alpha$, along with a dynamical phase diagram showing the regions of dominance of direct, exchange and origin saddles.

As $\eta\to 0$, the zeros form lines whose crossing of the real (resp. the imaginary) axis gives rise to a dynamical phase transition in the quantum-mechanical (resp. classical Brownian) problem, as in conventional many-body DQPTs~\cite{heylDynamicalQuantumPhase2018}. Employing the mapping of Fisher zeros onto point charges in 2D electrostatics~\cite{Fisher1965Nature,Fisher_equireview_2019}, we prove that the first- and second-order nonanalyticities in $g(\tau)$ correspond to lines of zeros of constant and linearly vanishing effective charge density respectively (see Appendix~\ref{app:zeros}). To our knowledge, such second-order nonanalyticities, caused by the crossing of two lines of Fisher zeros [see the vertical axis crossing in Fig.~\ref{fig:Fisher_zeros}(a)], have not been reported previously in many-body DQPTs.

\paragraph{Order parameter and optimal trajectories ---}The trajectory switch suggests a natural dynamical order parameter based on the trajectory endpoints. We consider a generating function:
\begin{align}
	M(\tau)=\int d\xi d\xi_0 \, \psi_0(\xi_0)K(\xi, \xi_0, \tau)\psi_0(\xi)e^{h\frac{\xi\xi_0}{\eta}}
\end{align}
and define an order parameter as $C(\tau)=\eta\partial_h\ln M(\tau)|_{h=0}$. For $\eta{\to}0$, $C(\tau)$ reduces to the weighted average, over the dominant saddles, of the trajectory endpoint products $\xi_0\xi$.

In the classical problem $C$ has a direct interpretation as an endpoint correlator. For the quantum case it takes the form of a so-called sequential weak value \cite{weak_pioneer, sequential_weak_pioneer},
\begin{align}\label{eq:weak_sequential}
C(\tau_x)=(\xi_{\tau_x}, \xi_0)_w=\frac{\braket{\psi_0|\hat \xi e^{-\frac{i}{\hbar\omega }\hat H\tau_x}\hat \xi|\psi_0}}{\braket{\psi_0|e^{-\frac{i}{\hbar\omega }\hat H\tau_x}|\psi_0}}.
\end{align}
The appearance of a weak value here is also natural in the view of earlier work using it as a tool for visualizing dominant semiclassical trajectories~\cite{TANAKA2002307, Turok_2014, nishimura2023newpicturequantumtunneling}. A related idea of using conditional expectation values for classifying DQPTs has been proposed~\cite{Canovi_firstorder} in the many-body context~\footnote{The resulting quantity in Ref.~\cite{Canovi_firstorder} was not identified as a weak value, but is essentially one. Unlike \eqref{eq:weak_sequential}, however, it is a weak value of a single observable rather than a weak correlator.}.

The measurement protocol in the quantum case is similar to the classical two-point construction: two \textit{weak} measurements of the particle coordinate should be performed at initial and final times for the evolution with preselection and postselection states given by $\ket{\psi_0}$. A sequential weak value is extracted from correlations between the two meter readouts rather than from a single pointer shift. Here the corresponding single-time weak values vanish, so the position-position correlation of the meters directly probes $\Re[C(\tau)]$. This quantity alone can serve as an order parameter~\cite{PhysRevA.76.044103}.

Figure~\ref{fig:order_parameter} shows $\Re[C(\tau_{x/y})]$ for several $\eta$ with $\alpha = 0.25$ and $\alpha = 0.8$ corresponding to panels $(a)$ and $(d)$ of Fig. \ref{fig:Fisher_zeros}. As $\eta$ decreases, the curves approach the asymptotic behavior given by the thick black lines. The order parameter has a discontinuity at the first-order quantum transition and is continuous at the classical second-order one. In both cases, order parameter changes coincide with the nonanalyticities of the SRF.

\paragraph{Discussion ---} We have shown that a single quantum particle can exhibit a DQPT in the semiclassical limit, and that analytic continuation of the problem describes a Brownian relaxation transition in the weak-noise limit. Both arise from the same complex-time Fisher-zero structure and admit the same trajectory-based order parameter. Thus, in single-particle systems, the weak-noise limit in the classical case and the semiclassical limit in the quantum case play essentially the same role. The mapping suggests that similar partner-problem constructions may extend the class of dynamical problems admitting an equilibrium-like description.

Even though our search for DQPT-like behavior in the relaxation problem was motivated by Ref.~\cite{meibohmFiniteTimeDynamicalPhase2022}, the phase transition considered in that work should be distinguished from the one discovered here. The nonanalyticity in the probability-distribution rate function for the Brownian particle in Ref.~\cite{meibohmFiniteTimeDynamicalPhase2022} and the one in the weighted-return rate function considered in this Letter share the same weak-noise mechanism of competing optimal trajectories, but correspond to different observables and occur in complementary parameter regimes. In the present parametrization, the transition in the probability-distribution rate function would occur for $\alpha>1/2$, whereas that of Eq. \eqref{eq:classical_SRF} exists for $\alpha<1/2$. Understanding more generally how different observables of the same system may host distinct dynamical phase transitions is an interesting subject for future studies.
\newpage
\bibliographystyle{unsrt}
\bibliography{references.bib}
\clearpage

\appendix
\setcounter{secnumdepth}{1}

\section{Construction of the confining potential for the quantum particle}\label{app:SUSY}

Having specified the desired ground state \eqref{eq:double-peak}, we introduce a superpotential $
W(x)=-d_x[\ln\psi_0(x)]=\frac{2\alpha m\omega}{\hbar a^2}x(x^2-a^2)$, using the terminology of supersymmetric QM, see e.g. \cite{Schwabl2007QM}. With the ground state energy chosen to be zero, the pre-quench potential ~$V_0(x)=\frac{\hbar^2}{2m}\left[W^2(x)-W'(x)\right]$ is
\begin{align}
	V_0(x)&=\frac{\alpha \hbar\omega}{a^2}\left[\frac{2\alpha m\omega}{\hbar a^2}x^2(x^2-a^2)^2-(3x^2-a^2)\right]
\end{align}

\section{Stationary points}\label{app:statpoints}
After a coordinate transformation $\xi_\pm=\frac{{\xi\pm\xi_0}}{\sqrt{2}}$,
Eq.~\eqref{eq:action} reads
\begin{align*}
	S(\xi_+, \xi_-){=}\frac{\alpha}4 (\xi_+^2{+}\xi_-^2{-}2)^2{+}\alpha \xi_+^2\xi_-^2+\frac{i}{2}\left[\xi_+^2\tan\left(\frac{\tau}{2}\right){-}\xi_-^2\cot\left(\frac{\tau}2\right)\right].
\end{align*}

The stationary points of $S$ are summarized in Table~\ref{tab:saddles}; their contributions to $G(\tau)$ are determined next.
\begin{table}[htp]
\centering
\renewcommand{\arraystretch}{1.5}
\begin{tabular}{c c c c}
\hline\hline
$\sigma$
&
$g_\sigma$
&
$(\xi_+^2,\xi_-^2)$
&
$S_\sigma$
\\
\hline

$0$
&
$1$
&
$(0,0)$
&
$\alpha$
\\

$d$
&
$2$
&
$\displaystyle
\left(
2-\frac{i}{\alpha}\tan\frac{\tau}{2},
\,0
\right)
$
&
$\displaystyle
\tan\frac{\tau}{2}
\left[
\frac{1}{4\alpha}\tan\frac{\tau}{2}+i
\right]
$
\\[3ex]

$e$
&
$2$
&
$\displaystyle
\left(
0,\,
2+\frac{i}{\alpha}\cot\frac{\tau}{2}
\right)
$
&
$\displaystyle
\cot\frac{\tau}{2}
\left[
\frac{1}{4\alpha}\cot\frac{\tau}{2}-i
\right]
$
\\[3ex]

$m$
&
$4$
&
$\displaystyle
\xi_\pm^2=
\frac{1}{2}
\pm
\frac{i}{4\alpha}
\frac{2\pm\cos\tau}{\sin\tau}
$
&
$\displaystyle
\frac{\alpha}{2}
-\frac{1}{8\alpha}\left(1+\csc^2\tau\right)
-\frac{i}{2}\cot\tau
$
\\[2ex]

\hline\hline
\end{tabular}
\caption{For each saddle type $\sigma$, we list its degeneracy $g_\sigma$, the
stationary-point coordinates, and the corresponding value $S_\sigma$.}
\label{tab:saddles}
\end{table}

\section{Survival amplitude saddle decomposition}\label{app:PL}

The contribution of the origin saddle can be fixed independently since it always remains on the original integration plane. For generic complex $\tau$ its
steepest-descent contour crosses $\mathbb R^2$ at the origin, so $n_0=1$.

For the others we use analytic continuation. The coefficients $n_\sigma$ are constant within \textit{Stokes chambers},
separated by the Stokes lines
$\ell^{\mathrm{Im}}_{\sigma\tilde{\sigma}}$ defined by
$\operatorname{Im}[S_\sigma{-}S_{\tilde{\sigma}}]=0$. We call a Stokes line \textit{active} if crossing it actually changes the
decomposition. Our strategy is to determine $n_\sigma$ where this is simple and then
analytically continue across the complex-$\tau$ plane. We use the following three \textbf{rules}, based on standard
Picard--Lefschetz theory~\cite{witten2010analyticcontinuationchernsimonstheory}
and direct inspection of $S(\xi_+,\xi_-)$:\\
\textbf{M:}
 When $S$ is real, nondegenerate \emph{minima} on the original
$\mathbb R^2$ integration plane contribute. Locally,
$S-S_\sigma=\frac12\sum_j\lambda_j z_j^2+\ldots$ with $\lambda_j>0$,
so $\Re [S]$ increases along real directions and decreases along imaginary
ones. Thus the steepest-descent directions cross the original plane
rather than run along it, and the contribution persists under a
sufficiently small complex deformation.\\
\textbf{T:} The Stokes transformations are \emph{triangular}: if
    $\Re [S_\sigma]<\Re [S_{\tilde\sigma}]$ at the crossing of a Stokes line, the coefficient $n_\sigma$
    of the more dominant saddle is unchanged, while
    $n_{\tilde\sigma}\to
    n_{\tilde\sigma}-m_{\sigma\tilde\sigma}n_\sigma$,
    with $m_{\sigma\tilde\sigma}\in\mathbb Z$. If $m_{\sigma\tilde\sigma}=0$, the Stokes line is inactive.\\
\textbf{G:} A \emph{global} criterion $\Re [S_\sigma]{<}\min\limits_{\mathbb R^2,\;\Im [S]=\Im [S_\sigma]}\Re [S]$ excludes a saddle $\sigma$.
As in the familiar 1D steepest-descent picture, $\Im [S]$
is constant along a steepest path while $\Re [S]$ changes monotonically.
Thus, a steepest path starting from $\sigma$ cannot reach the original plane $\mathbb R^2$. In our case, the fixed-phase condition is linear in $\xi_\pm^2$, whereas
$\Re [S]$ is quadratic, so the constrained minimum is obtained analytically.

\begin{figure}
    \centering
    \hspace*{-1cm}
    \includegraphics[trim = 0mm 0mm 0mm 0mm,
clip,width=0.9\linewidth]{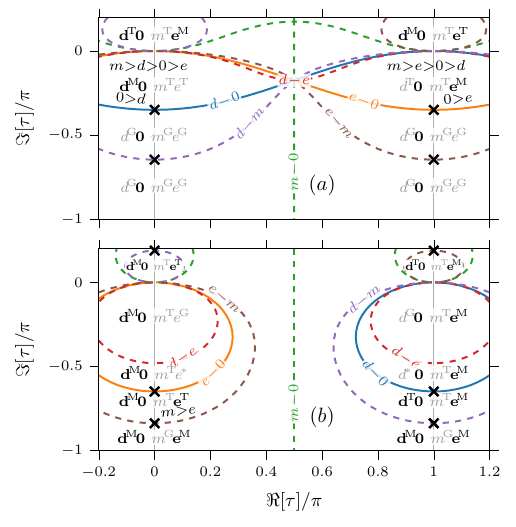}
    \caption{
Stokes lines for (a) $\alpha=0.25$ and (b) $\alpha=0.65$. Solid (dashed) lines are active (inactive); crosses mark coalescences. Chamber labels list $d\,0\,m\,e$, with bold black (gray) symbols denoting
present (absent) saddles; superscripts refer to the criteria defined in the text.
Pairwise dominance needed for applying $T$ is indicated explicitly
(e.g. $d>0$) and evaluated with an infinitesimal $\epsilon$-shift.
}
    \label{fig:Stokes}
\end{figure}

We apply these rules along $\tau = \epsilon + \pi k-i\tau_y$,$\epsilon\to0$, starting at large $\tau_y$ and continuing toward the origin. The results are given in Fig.~\ref{fig:Stokes} for two representative cases, $\alpha=0.25<0.5$ and $\alpha=0.65>0.5$. The superscripts indicate one sufficient justification for each saddle assignment. Only once do we need an additional argument: when crossing the
$e$--$0$ ($d$--$0$) Stokes line for even (odd) $k$ when $\alpha>0.5$,
see the asterisk superscripts in Fig.~\ref{fig:Stokes}. There is,
however, a saddle coalescence on these lines, which simplifies the
analysis. The two saddles lie on an invariant coordinate axis, where
locally $S-S_0=\frac{\mu}{2}z^2+\frac{\alpha}{4}z^4$ with $z=\xi_-$ ($\xi_+$) for even (odd) $k$ and $\mu=0$ at the
coalescence. The transverse quadratic mode remains nondegenerate,
$h_\perp=(1-4\alpha^2)/(2\alpha)\neq0$, so the local problem reduces to
the familiar 1D quartic steepest-descent problem and fixes the remaining
saddle assignment.

\begin{table*}[htp]
\centering
\renewcommand{\arraystretch}{1.5}
\setlength{\tabcolsep}{6pt}
\begin{tabular}{c c c c c}
\hline\hline
Regime
&
$\{\sigma,\tilde{\sigma}\}$
&
Critical time
&
Leading $\Delta S_{\sigma\tilde{\sigma}}$
&
Fisher-zero density
\\
\hline

Quantum, $\alpha<1/2$
&
$\{d,0\}$
&
$\displaystyle \tau_x^*=2\arctan(2\alpha)$
&
$\displaystyle
2i\alpha
+(1+i)\left(\frac12+2\alpha^2\right)\delta\tau
+O(\delta\tau^2)
$
&
$\displaystyle
\lambda(0)=
\frac{1+4\alpha^2}{2\sqrt{2}\pi}
$
\\

Quantum, $\alpha>1/2$
&
$\{d,e\}$
&
$\displaystyle \tau_x^*=\frac{\pi}{2}$
&
$\displaystyle
2i+\frac{\delta\tau}{\alpha}
+O(\delta\tau^2)
$
&
$\displaystyle
\lambda(0)=\frac{1}{2\pi\alpha}
$
\\

Classical, $\alpha<1/2$
&
$\{d,0\}$
&
$\displaystyle
\tau_y^*=2\operatorname{arctanh}(2\alpha)
$
&
$\displaystyle
\frac{(4\alpha^2-1)^2}{16\alpha}\,
\delta\tau^2
+O(\delta\tau^3)
$
&
$\displaystyle
\lambda(s)=
\frac{(4\alpha^2-1)^2}{16\pi\alpha}|s|
+O(s^2)
$
\\[0.8ex]

\hline\hline
\end{tabular}
\caption{Leading order $\Delta S_{\sigma\tilde\sigma}$ and Fisher-zero densities
near representative crossings. Periodic
repetitions have the same local
structure.}
\label{tab:fisher_zeros}
\end{table*}

\begin{figure}[htp]
    \centering
    \hspace*{-1cm}
    \includegraphics[trim = 2mm 6mm 2mm 6mm,
clip,width=.9\linewidth]{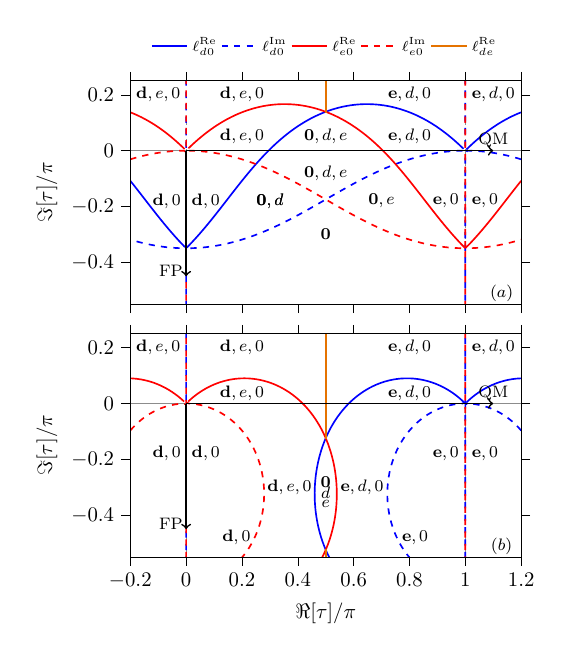}
    \caption{The active Stokes (dashed) and anti-Stokes (solid) lines for (a) $\alpha=0.25$ and (b) $\alpha = 0.65$. In each chamber the labels of active saddles are listed with the dominant one in bold font.}
    \label{fig:statfiz}
\end{figure}

\paragraph{Numerical sampling}

In Fig.~\ref{fig:3Dtraj} we visualize finite-$\eta$ fluctuations around the contributing saddles $\mathbf z_\sigma=(\xi_{0,\sigma},\xi_\sigma)$. Writing $z= z_\sigma+\sum_a u_a\bs e_a^{(\sigma)}$, where $\mathbf e_a^{(\sigma)}$ are the two local steepest-descent directions, the action for sufficiently small deviations takes the quadratic form
\begin{equation}
    S(\mathbf z)\simeq S_\sigma+\frac{1}{2}\sum_a \lambda_{\sigma a}u_a^2,\qquad \lambda_{\sigma a}>0.
\end{equation}
Then we sample the corresponding Gaussian fluctuations. Their widths scale as $\sqrt{\eta/\lambda_{\sigma a}}$, so unequal local curvatures produce the elongated distributions seen in the figure. In Fig.~\ref{fig:2Dtraj} we directly sample Eq.~\eqref{eq:FP}, starting from $P_0$, and retain trajectories with endpoint probability proportional to $P_0(\xi_\tau)/P_{\rm eq}(\xi_\tau)$. Their histogram thus samples the weighted return in Eq.~\eqref{eq:g_FP}.

\section{Locating Fisher zeros}\label{app:zeros}

Once the contributing saddles are identified, Fisher zeros arise from
destructive interference between two competing saddle contributions
$\sigma$ and $\tilde\sigma$. Neglecting subleading prefactors, their
positions are determined by ($\Delta S_{\sigma\tilde\sigma}=S_{\sigma}-S_{\tilde \sigma}$):
\begin{equation}
    \Re[\Delta S_{\sigma\tilde\sigma}]=0,\quad
    \Im[\Delta S_{\sigma\tilde\sigma}]=(2k+1)\pi\eta.
\end{equation}
The first condition defines the equal-weight line, while the second
fixes the zeros along it, with zeros occurring only on those segments
where both saddles contribute to the integration cycle. Each zero carries an ``effective charge'' $\eta$. The corresponding asymptotic line density is
\begin{align}
    \lambda(s)
    =\lim_{\eta\to0}\eta\frac{dk}{ds}
    =\frac{1}{2\pi}
    \left|\frac{d}{ds}\Im[\Delta S]\right|
\end{align}
where $s$ parametrizes the equal-weight line.

The three relevant crossings are summarized in
Table~\ref{tab:fisher_zeros}. For the quantum transitions on the real-time axis, $\Delta S$ is
linear in $\delta\tau$, and the Fisher-zero lines reach the axis with
finite density. For the $d$--$0$ transition at $\alpha<1/2$, the
zero line crosses the real-time axis at angle $\pi/4$, yielding a jump
in the first derivative, $\left|\Delta g'(\tau_x^*)\right|
=2\pi\lambda(0)\cos({\pi}/{4}).$
For the $d$--$e$ switch at $\alpha>1/2$, the zero lines periodically
cross the real-time axis perpendicularly, and $\left|\Delta g'(\tau_x^*)\right|
=2\pi\lambda(0).$

The classical transition is qualitatively different. At
$\tau_y^*=2\operatorname{arctanh}(2\alpha)$, the direct saddles merge
with the origin, so the linear term in $\Delta S_{d0}$ vanishes.
The equal-weight condition then produces two lines,
$\Re\delta\tau=\pm\Im\delta\tau$, each meeting the imaginary-time axis
at angle $\pi/4$. The zero density vanishes linearly along the relevant
segments. Correspondingly,
$\left|
\Delta\partial_{\tau_y}^2 g(-i\tau_y^*)
\right|
=
2\pi\lim_{s\to0}\frac{\lambda(s)}{|s|}.$

Thus, in analogy with the equilibrium Fisher-zero picture, the
first-order quantum nonanalyticities are associated with lines of
finite density of zeros, whereas at the second-order classical transition
the density vanishes linearly at the critical point.

\end{document}